# The Effect of 45° Grain Boundaries and associated Fe particles on $J_c$ and resistivity in Ba(Fe$_{0.9}$Co$_{0.1}$)$_2$As$_2$ Thin Films

J. Hänisch[a+], K. Iida[a], F. Kurth[a,b], T. Thersleff[a*], S. Trommler[a,b], E. Reich[a], R. Hühne[a], L. Schultz[a,b], and B. Holzapfel[a,c]

[a]IFW Dresden, Helmholtzstr. 20, 01069 Dresden, Germany
[b]TU Dresden, Dept. Mathematics and Natural Science, 01062 Dresden, Germany
[c]TU Bergakademie Freiberg, Akademiestr. 6, 09596 Freiberg, Germany
*Now at: The Ångström Laboratory, Uppsala University, 752 37 Uppsala, Sweden
+Email: j.haenisch@ifw-dresden.de

**Abstract.** The anisotropy of the critical current density $J_c$ depends in general on both the properties of the flux lines (such as line tension, coherence length and penetration depth) and the properties of the defects (such as density, shape, orientation etc.). Whereas the $J_c$ anisotropy in microstructurally clean films can be scaled to an effective magnetic field containing the Ginzburg-Landau anisotropy term, it is in general not possible (or only in a limited field range) for samples containing extended defects. Here, the $J_c$ anisotropy of a Co-doped BaFe$_2$As$_2$ sample with 45° [001] tilt grain boundaries (GBs), i.e. grain boundaries created by 45° in-plane rotated grains, as well as extended Fe particles is investigated. This microstructure leads to $c$-axis correlated pinning, both due to the GBs and the Fe particles and manifests in a $c$-axis peak in the $J_c$ anisotropy at low magnetic fields and a deviation from the anisotropic Ginzburg-Landau scaling at higher fields. Strong pinning at ellipsoidal extended defects, i.e. the Fe particles, is discussed, and the full $J_c$ anisotropy is fitted successfully with the vortex path model. The results are compared to a sample without GBs and Fe particles. 45° GBs seem to be good pinning centers rather than detrimental to current flow.

## INTRODUCTION

The iron-based superconductors of the so-called 122 family with ThCr$_2$Si$_2$ type structure,[1] e.g. SrFe$_2$As$_2$ (Sr-122) and BaFe$_2$As$_2$ (Ba-122) are the most investigated pnictides in thin film form because of their relatively high critical temperature $T_c$ (compared to the 11 family, e.g. FeSe and FeSe$_{0.5}$Te$_{0.5}$) and their ease of preparation (compared to the 1111 family, e.g. LaFeAsOF, SmFeAsOF, where *ex-situ* phase formation [2] or molecular beam epitaxy [3] is necessary). The high upper critical field with its low anisotropy (~1) at low temperature, $T$, and the low Ginzburg number ($Gi \sim 10^{-4}$) [4] let the 122 compounds appear interesting for applications.

There are three possibilities for providing the charge carrier density necessary for superconductivity in the 122 compounds: i) electron doping by substitution of Fe with e.g. Co, ii) hole doping by substitution of the alkaline element with e.g. K, and ii) applying chemical pressure via isovalent doping by e.g. P substitution at the As site [5,6] or water-intercalation.[7] Because of the low vapor pressure of Co and its easy incorporation in the 122 crystal structure, especially Co-doped Ba-122 has been investigated widely, even though maximum $T_c$ of K-doped compounds is appreciably higher.[1]

Due to the low charge carrier density and the d-wave order parameter symmetry in high-$T_c$ superconducting cuprates, grain boundaries (GBs) have a detrimental effect on $J_c$ in these compounds. Recently, it was shown that Ba-122 shows the same exponential drop in $J_c$ with GB angle but with a slope three times smaller [8,9]. The smaller slope might be due to the more isotropic order parameter symmetry (s++ and s± are discussed) and the lower $Gi$ number. Weiss et al. have shown that by adjusting the preparation conditions, the GB issue can be eliminated in powder-in-tube Ba-122 tapes,[10] which is very promising for potential applications.

After $J_c$ scaling with the Anisotropic Ginzburg-Landau description, as discussed below, in microstructurally clean Co-doped Ba-122 thin films had been found,[11] a Co-doped Ba-122 thin film with increased defect density, namely GBs and related, sparsely distributed Fe precipitates, shall be discussed in this contribution.

## EXPERIMENTAL

The Co-doped Ba-122 thin film of this study was prepared by pulsed laser deposition (PLD) under UHV condition from a stoichiometric sinter target of composition Ba(Fe$_{0.9}$Co$_{0.1}$)$_2$As$_2$ with a KrF excimer laser ($\lambda$ = 248 nm) operated at a repetition rate of $f$ = 10 Hz and an energy density at the target surface of $E$ = 5 mJ/cm$^2$. The substrate was a (001) oriented (La,Sr)(Al,Ta)O$_3$ (LSAT) single crystal of 10 × 10 × 1 mm$^3$, heated to 675 °C during deposition. A growth rate of ~0.1 Å/pulse and a pulse number of 8000 resulted in a thickness of 75 nm, determined by transmission electron microscopy (TEM). More details of the film preparation can be found in Ref. [12]. Structural properties of the sample were investigated by means of x-ray diffraction (XRD) in θ-2θ geometry at a Bruker D8 Advance with Co K$_\alpha$ radiation and at a texture goniometer Phillips X'pert with Cu K$_\alpha$ radiation, as well as TEM on an FEI Titan operated at 300 kV with C$_s$ correction for parallel illumination. Electrical transport properties were measured in four-point geometry in a PPMS (Quantum Design) applying magnetic fields up to 9 T on bridges of 1 mm length and 250 µm width, prepared by Ar ion milling. From $R(B)$ and $R(T)$ curves, the upper critical field, $H_{c2}$, and the irreversibility field, $H_{irr}$, were determined with a criterion of 90% $R(25\ \text{K})$ (i.e. the resistance at onset $T_c$) and 1% $R(25\ \text{K})$, respectively. $J_c$ was determined from $V(I)$ curves with an electric field criterion of 1 µV/cm.

## STRUCTURAL PROPERTIES

The Co-doped Ba-122 film grew mainly *c*-axis oriented as revealed in the θ-2θ scan, Fig. 1(a). Only (00*l*) peaks of Ba-122, besides a tiny Ba-122 (110) peak at 38°, and the substrate peaks are seen. Due to the relatively low deposition temperature, the in-plane texture of this film is not perfect but shows 45° in-plane rotated grains of 3.3 % volume fraction, Fig. 1(b). The texture of the main, epitaxial phase is very sharp though, with full-width at half-maximum (FWHM) values (not corrected for device broadening, ~0.3°) of $\Delta\phi$ = 1.18° for the (103) reflection and $\Delta\omega$ = 0.45° of the (004) rocking curve, Fig. 1(c). Due to the 45° in-plane rotated grains, [001] tilt GBs have developed in the Ba-122 layer. As seen in TEM images, Fig. 2, these GBs are mostly straight throughout the Co-doped Ba-122 layer and often contain precipitates within their structure or nearby. By analyzing Moiré fringes, it was deduced that these particles consist of Fe with orientation relationship (001)[110]Fe∥(001)[100]Ba-122, and indeed, at a closer look at the θ-2θ scan, a tiny peak at 77° can be made out, which refers to Fe (002). It was observed in previous films on bare oxide substrates that, under certain conditions, a thin interfacial Fe layer of around 2 nm thickness forms between the substrate and the Co-doped Ba-122 film.[11] The formation of the Fe precipitates might be caused by a similar mechanism: a higher desorption rate for As at the beginning of film growth due to the lack of the stable 122 compound leads to the formation of elemental Fe and probably diffusion of Ba into the substrate.[11] The higher mobility in GBs and the disturbed 122 crystal structure may also lead to higher As desorption and similar formation of Fe particles. Due to the low misfit between Fe(001) and the Fe basal plane of Ba-122,[13] these particles grow biaxially textured.

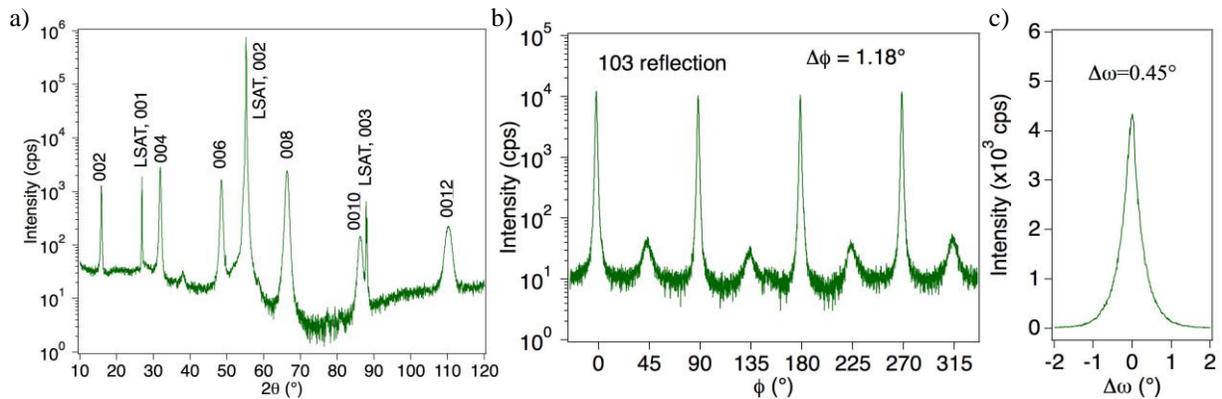

**FIGURE 1.** XRD investigation of the Co-doped Ba-122/LSAT film. a) θ-2θ scan at χ = 90°, b) φ scan at χ = 41.4° [the (103) reflection], c) ω scan (rocking curve) for the (004) reflection.

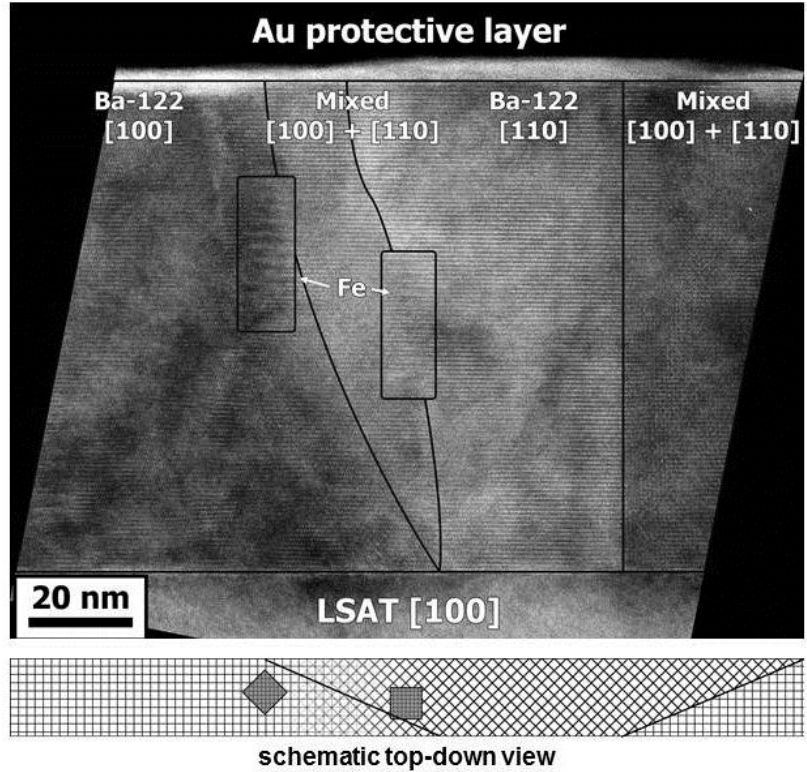

**FIGURE 2.** Cross sectional TEM image of the Co-doped Ba-122/LSAT film, revealing a sharp substrate-film interface, 45° [001] tilt GBs (vertical black lines) and round and *c*-axis elongated Fe precipitates (encircled) near or within the GBs. The image was taken in the (100) pole. Below the TEM image is a schematic top-down view illustrating the orientation of GBs and Fe particles, deduced from Moiré contrasts.

# ELECTRICAL TRANSPORT PROPERTIES

## Resistance and Superconducting Transition

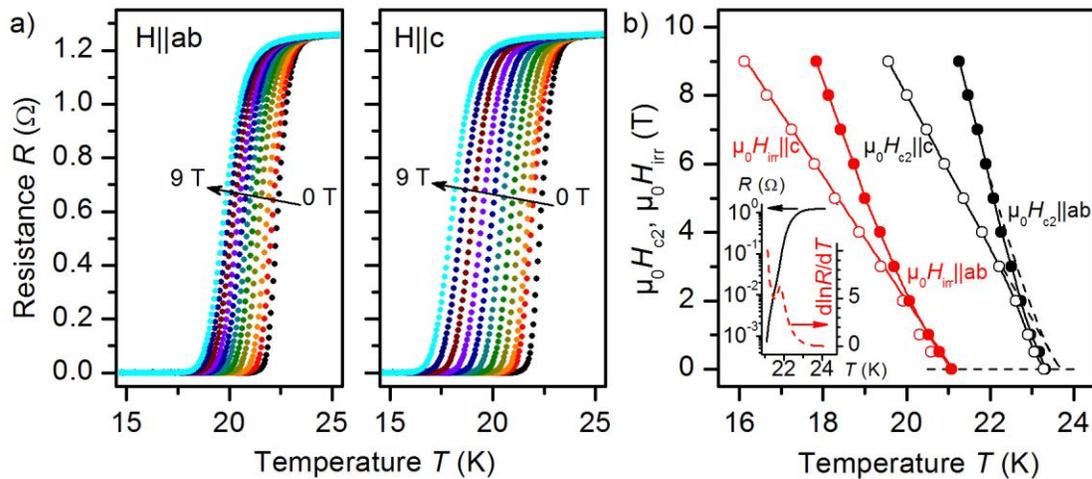

**FIGURE 3.** a) Resistive transition of the Co-doped Ba-122 film for both major directions, **H||ab** and **H||c**, between 0 T and 9 T in 1 T steps, additionally 0.5 T, b) temperature dependence of $H_{c2}$ and $H_{irr}$ for both major directions. Inset: *R(T)* in zero field in log. scale and $d\ln R/dT$ revealing a second transition.

Figure 3(a) shows the resistive transitions for magnetic fields applied in both major directions, **H||ab**, and **H||c**. The latter direction exhibits nearly parallel transitions with applied fields, indicating a linear dependence of both $H_{c2}$ and $H_{irr}$ on temperature, Fig. 3(b). $H_{c2}$ shows a change in curvature near $T_c$ for both directions. This behavior is due to the reduced $T_c$ in the GB regions. Firstly, the zero-field transition and its log. derivative [inset Fig. 3(b)] show a small foot structure, known from cuprates with GBs.[14] Secondly, if the high-field $H_{c2}$ data are extrapolated with the same functions as found for $H_{irr}$ [i.e. $H_{irr} \sim (1-T/T_c)^q$, with $q = 1$ and 1.3 for $c$-direction and $ab$-plane respectively] towards $\mu_0 H = 0$ T (dashed lines), the shift in $T_c$ of around 0.4 K corresponds to the difference in the $T_c$ values from $R(T, 0\,\text{T})$, inset Fig. 3(b). $H_{irr}^c = H_{irr}||c$ is also slightly reduced for fields below 2-3 T, probably due to vortex channeling effects. It is assumed but still to be confirmed that this field corresponds to the cross-over field between inter- and intragrain $J_c$ limitation.[15]

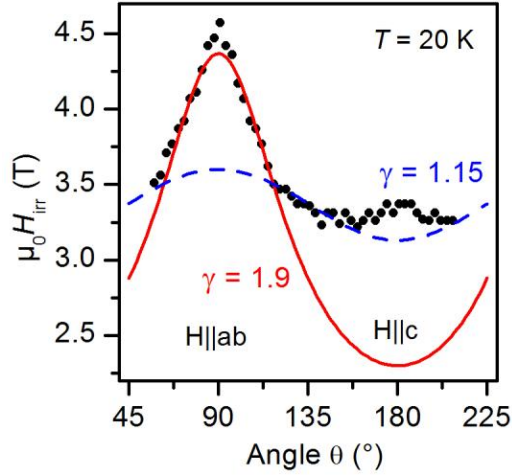

**FIGURE 4.** Angular dependence of $H_{irr}$ at 20 K. The lines are fits to Eq. 1 with $\gamma = 1.9$ (full) and 1.15 (dashed), respectively.

The full anisotropy of the irreversibility field $H_{irr}$ at 20 K, Fig. 4, was determined from $R(H)$ data for several angles θ (angle between applied magnetic field and crystallographic $c$-axis) in the same way as $H_{irr}(T)$. Three angular regions are visible: Around the $ab$-direction, $H_{irr}$ is well described by the anisotropic Ginzburg-Landau equation

$$\frac{H_{irr}(\theta)}{H_{irr}^c} = 1/\sqrt{\cos^2(\theta) + \gamma^{-2}\sin^2(\theta)} \qquad (1)$$

The anisotropy parameter γ is here 1.9 and corresponds well with the anisotropy of the upper critical field at this temperature. In the absence of correlated defects, $H_{c2}$ and $H_{irr}$ are expected to show the same anisotropy behavior. Around $c$, there is a second peak in $H_{irr}$ visible, most likely due to correlated pinning at the GBs. In the region between these two peaks, the $H_{irr}$ data can be fitted with Eq. 1 as well, however here with a much lower (effective) γ = 1.15. The reason for that lies in the presence of small but extended Fe particles. Recently, van der Beek *et al.* showed theoretically that strong pinning centers slightly larger than the in-plane coherence length can lead to a reduction of the apparent anisotropy in uniaxially anisotropic superconductors.[16] The deviation at **H||ab** may be explained by surface pinning and occasional $ab$-planar defects.

## Critical Current Density

Figure 5 shows the angular dependence of the critical current density, $J_c(\theta)$, for three medium temperatures 12 K, 14 K, and 16 K, under maximum Lorentz force configuration, i.e. **J** ⊥ **H** at all angles θ. The lower panels show the same data normalized to $J_c$ at θ = 90° (**H||ab**). This sample not only shows the $J_c$ peaks for **H||ab** due to electronic anisotropy and surface pinning, but also a pronounced $c$-axis peak (θ = 180°) as expected from the angular dependence of $H_{irr}$ (Fig. 4). There are two reasons for this $c$-axis peak. Firstly, the GBs are predominantly oriented perpendicular to the $ab$-plane and, hence, act as $c$-axis correlated pinning centers. Secondly, it was shown recently by van der Beek *et al.* [16] theoretically and, e.g., by Yamasaki *et al.* experimentally [17] that any strong defect extended in its dimensions beyond the respective coherence length will contribute anisotropically to $J_c$, in this case enhancing the *c*-axis share. This contribution increases with the elongation of these particles in *c*-axis direction. The Fe particles found in our film have a *c:a* aspect ratio of 1:5, and are, therefore, expected to contribute to the *c*-axis peak as well. The peak intensity, relative to the *ab* peak, is maximal at an applied field of around 2 T, irrespective of temperature *T*. This field seems to correspond to the density of GBs and extended Fe particles. For a certain applied magnetic field, the (relative) *c*-axis peak is decreasing with decreasing *T*, see e.g. the 2 T data in the lower panel. This effect is well known from $YBa_2Cu_3O_{7-d}$ (YBCO) based thin films and coated conductors and is most likely caused by a faster increase of the *ab*-peak than the *c*-axis peak with decreasing *T*, which is ultimately linked to both angular and temperature dependence of the shear modulus $C_{66}$ of the flux line lattice.

Some curves show a slight shift of the *c*-axis peak by 1-3°. This is probably caused by an asymmetric distribution of GB directions in the measurement bridge (*cf.* the tilted GBs in Fig. 2). The *c*-axis peak of this sample cannot be attributed to correlated pinning (i.e. GBs) or ellipsoidal strong defects (i.e. Fe particles) alone. In the first case, the width of this peak should decrease with increasing applied magnetic field, but is rather constant. And no conclusive fits with the model for strong elliptical defects [16] was possible for reasonable fit parameters for particle size and anisotropy of coherence length and penetration depth. Therefore, the contribution of both types of defects overlay at the *c*-axis peak.

The bicrystal experiments of Katase *et al.* [8] suggest that the critical current densities across GBs with angle α = 45° lie significantly above the exponential $J_c(\alpha)$ dependence found for all other GB angles α. This effect, never observed for cuprates, implies that a 45° in-plane rotation in Ba-122 may lead to a special GB with narrow interface and clean interfacial structure. These GBs seem to be effective pinning centers already at fields around 2 T rather than detrimental to the current flow. Further TEM and transport studies have to validate this hypothesis.

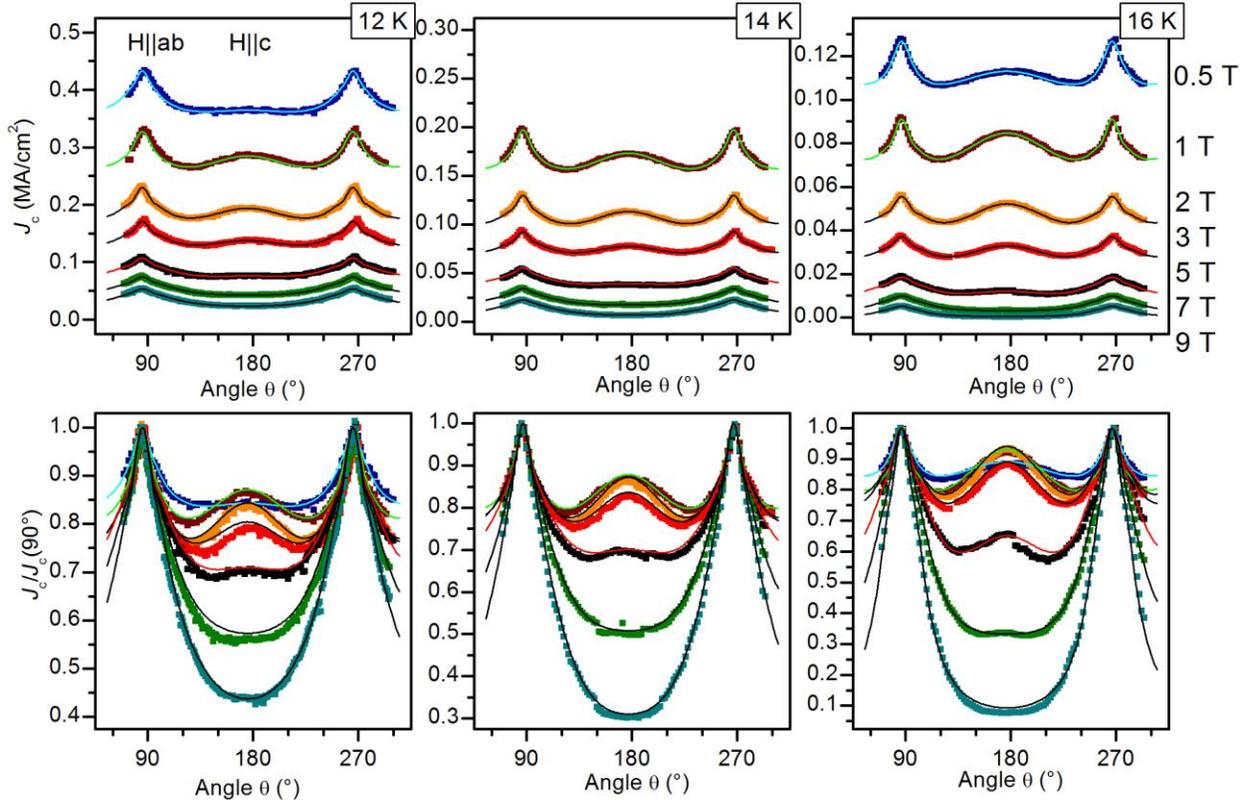

**FIGURE 5.** Angular dependence of $J_c$ at 12 K, 14 K, and 16 K. The upper panels show absolute values, the lower panels show the same data normalized to the $J_c$ value at 90°. The lines are fits to the vortex path model.

In the absence of correlated disorder, the anisotropy of the upper critical field and the irreversibility field are the same (Eq. 1) and are reflecting in the $J_c$ anisotropy. This can be shown by scaling the angular dependent $J_c$ data to an effective magnetic field $H_{eff}$ according to [18,19]:

$$H_{eff} = H\varepsilon(\theta) \quad \text{with} \quad \varepsilon(\theta) = \sqrt{\cos^2\theta + \gamma^{-2}\sin^2\theta} \tag{2}$$

This is shown in Fig. 6 at $T = 12$ K for the sample of this study and a microstructurally clean sample [11]. Whereas the latter sample shows $J_c$ scaling in a wide angular range around $\mathbf{H}\|\mathbf{c}$, the sample containing GBs and Fe particles shows the scaling only in a limited angular range. Around the *c*-direction, the scaling deviates (as for the *ab*-direction for both samples). However, scaling parameter $\gamma = \gamma_{Hc2}(12\text{ K})$ as well as the envelope function for $J_c(H_{eff})$ (lines in Fig. 6, derived from Kramer fits to the pinning force density) are the same for both samples. Therefore, we can conclude that the same pinning centers as well as pinning mechanism is active in the scaling regions for both samples. Furthermore, even though there is no *c*-axis peak visible for 9 T (Fig. 5), the 9 T curve still clearly deviates from the scaling curve for random point-like disorder. This illustrates that the absence of a *c*-axis peak is not sufficient evidence for the absence or inactivity of extended or correlated defects. Recent high-field measurements showed that at fields above around 15 T the effect of such defects does disappear, resulting in a $J_c(H_{eff})$ matching the scaling law.[20]

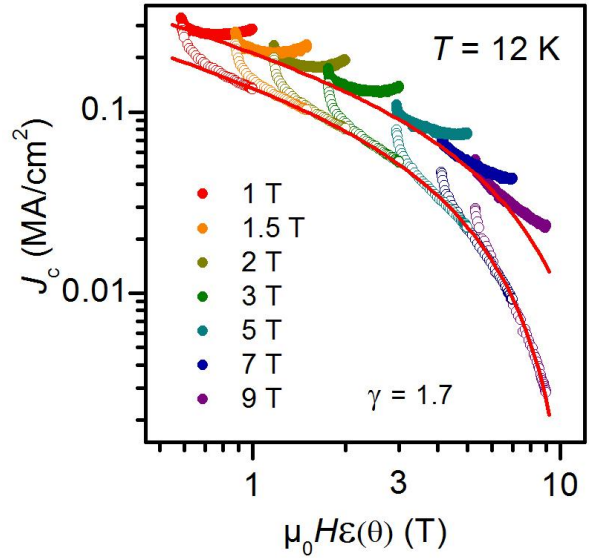

**FIGURE 6.** Scaling of $J_c$ to an effective magnetic field according to Eq. 2 for a microstructurally clean sample [11] (open symbols) and the sample with GBs and Fe particles of this study (full symbols). The lines are Kramer fits to the (effective) pinning force density $f_{p,eff} = J_c * \mu_0 H_{eff}$.

In general, the critical current density, and especially its anisotropy, depends both on extrinsic parameters, i.e. the pinning landscape, and on intrinsic parameters, such as coherence length and penetration depth as well as their temperature, field and angle dependence. Most theoretical models can only handle one kind of defect: e.g. the scaling approach describes point like random disorder, and ref. [16] describes strong ellipsoidal nanoparticles. Therefore they are not sufficient to describe the full $J_c(\theta)$ curves completely, also because the relation for the summation of pinning forces arising from different kinds of defects is generally unknown. Long *et al.* [21,22] developed therefore a model purely based on statistical considerations and the principle of maximum entropy. According to this model ("vortex path model", VPM), there exist certain "vortex paths", i.e. paths through the sample thickness with effective pinning at given *T* and *H*. These paths combine the anisotropies of the pinning centers and of the vortices and have a certain angular distribution with maximum entropy (i.e. Gaussian or Lorentzian shaped), which result in related $J_c(\theta)$ dependencies. The data of this study's sample could be fitted with a combination of Lorentzian contribution for the *ab*-peak and Gaussian contribution for the *c*-axis peak, shown as lines in Fig. 5. A detailed analysis of the peak shapes is to be published elsewhere. Whether there is a reasonable link between a constant background, $J_0$, (itself derived from a Lorentzian contribution) sometimes necessary in the VPM and the envelope in the scaling approach is still under discussion and has to be part of future studies. Evidence for such a possible link is the fact that $J_0$ and the envelope of the scaling approach show the same magnetic field dependence (not shown here).

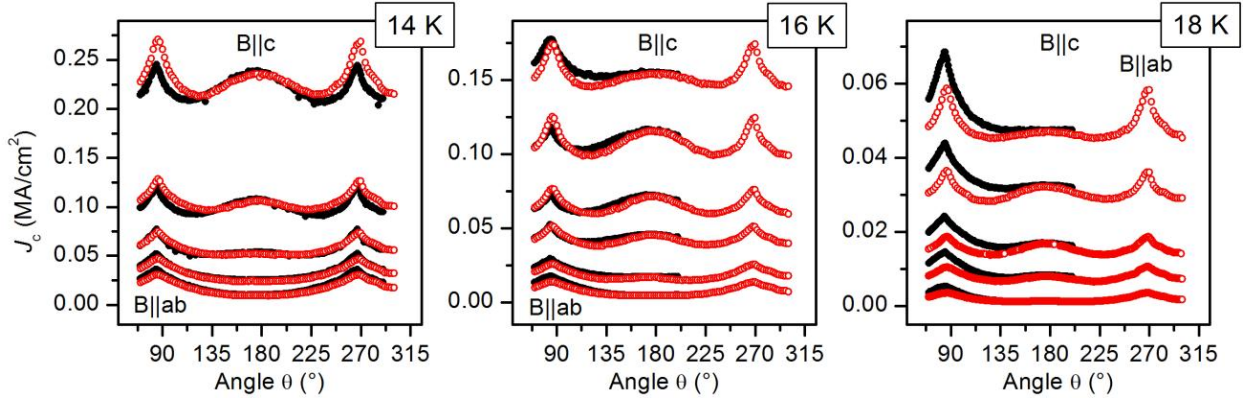

**FIGURE 7.** Variable Lorentz force measurements (full symbols) in comparison to maximum Lorentz force measurements (open symbols) of $J_c(\theta)$ for three temperatures and several fields, 14 K: 1-9T in 2 T steps, 16 K: additionally 0.5 T, 18 K: 1-5T in 2 T steps as well as 0.5 T and 2 T.

Additionally to the measurements under maximum Lorentz force configuration (MLF), measurements under variable Lorentz force (VLF, here: axis of field rotation $\perp c$-axis and $\perp \mathbf{J}$) were carried out. In this case, $\theta$ corresponds to the angle between field and current direction. This is shown for three representative temperatures in Fig. 7. At 18 K, $J_c(B//ab)$ is always larger for VLF than for MLF. This is expected in the case that the Lorentz force is determining $J_c$. However, in our sample, there seems to be a crossover around 16 K, where both configurations yield very similar curves and $J_c$ is, therefore, only dependent on the $c$-axis component of the applied magnetic field. At even lower $T$, the behavior may be reversed; now the MLF can lead to higher $J_c$ values for small applied fields. The effect of configuration independent $J_c(\theta)$ curves has recently been measured for a wide temperature range at a sample with a large density of $ab$-planar defects.[23] It may, therefore, be caused by different $J_c(T)$ dependences of different kinds of defects. This has to be cross-checked with similar measurements on microstructurally clean samples. The $c$-axis peaks in both configurations are very similar at all temperatures. This effect has been found previously by Maiorov *et al.* on YBCO thin films.[24] Another possible reason is a large microscopic meandering of the current flow and a smearing-out of the Lorentz force. This is plausible because Ba-122 is a 3D superconductor with small anisotropy. A similar effect has been observed by Rutter *et al.* on the in-plane VLF (i.e. field rotation axis equals $c$-axis) in YBCO coated conductor samples.[25]

## CONCLUSION

This study investigated the influence of 45° [001] tilt GBs and concomitant Fe particles in PLD-grown Co-doped Ba-122 thin films on the resistive transition and the angular dependence of $J_c$. GBs and Fe particles were found in XRD and TEM measurements. The GBs seem to broaden the transition in applied fields below 3 T, which might correspond to the crossover field between inter and intra-grain $J_c$ limitation. The $J_c(\theta)$ dependence shows a broad $c$-axis peak, which is maximal for applied fields around 2 T and is decreasing (with respect to the $ab$-peak) with decreasing applied field. This peak is caused by both GBs and nanoparticles. In a small angular range at medium angles, scaling of $J_c$ to the $H_{c2}$ anisotropy is found, similar to microstructurally clean films. A full description of $J_c(\theta)$ is possible with the vortex path model. VLF measurements suggest a wide $T$ range, where $J_c$ only depends on the magnetic field's $c$-axis component.

## ACKNOWLEDGMENTS

The authors thank G. Fuchs, S. Haindl, S. Wimbush and C. J. van der Beek for fruitful discussions, M. Sparing and R. Gärtner for help with the gold electrode preparation as well as J. Scheiter for help with FIB cut samples. We are also grateful to M. Kühnel and U. Besold for their technical support. This work was partially supported by the EU Marie-Curie RTN NESPA, the German Research Foundation, DFG, and the Leibniz Association.